%% file: main.tex
\documentclass[10pt, conference]{IEEEtran}
\IEEEoverridecommandlockouts
\usepackage{cite}
\usepackage{amsmath,amssymb,amsfonts}
\usepackage{algorithmic}
\usepackage{graphicx}
\usepackage{textcomp}
\usepackage{xcolor}
\usepackage{url}
\usepackage{enumitem}
\usepackage{tikz}

\definecolor{mygreen}{rgb}{0.0, 0.5, 0.0}

\newcommand{\ea}{\textit{et al.}}

\newcommand{\smallsection}[1]{\noindent {\bf #1}.\hspace{1mm}}

\newcommand{\rqone}{How accurate is ChatGPT for function and line-level vulnerability predictions?}
\newcommand{\rqtwo}{How accurate is ChatGPT for vulnerability types classification?}
\newcommand{\rqthree}{How accurate is ChatGPT for vulnerability severity estimation?}
\newcommand{\rqfour}{How accurate is ChatGPT for automated vulnerability repair?}

\def\BibTeX{{\rm B\kern-.05em{\sc i\kern-.025em b}\kern-.08em
    T\kern-.1667em\lower.7ex\hbox{E}\kern-.125emX}}

\begin{document}

\title{ChatGPT for Vulnerability Detection, Classification, and Repair: How Far Are We?}


\author{\IEEEauthorblockN{Michael Fu}
\IEEEauthorblockA{
\textit{Monash University}\\
Clayton, Australia \\
yeh.fu@monash.edu}
\and
\IEEEauthorblockN{Chakkrit (Kla) Tantithamthavorn}
\IEEEauthorblockA{
\textit{Monash University}\\
Clayton, Australia \\
chakkrit@monash.edu
}
\and
\IEEEauthorblockN{Van Nguyen}
\IEEEauthorblockA{
\textit{Monash University}\\
Clayton, Australia \\
Van.Nguyen1@monash.edu
}
\and
\IEEEauthorblockN{Trung Le}
\IEEEauthorblockA{
\textit{Monash University}\\
Clayton, Australia \\
trunglm@monash.edu
}
}


\maketitle

\begin{abstract}
Large language models (LLMs) like ChatGPT (i.e., gpt-3.5-turbo and gpt-4) exhibited remarkable advancement in a range of software engineering tasks associated with source code such as code review and code generation. In this paper, we undertake a comprehensive study by instructing ChatGPT for four prevalent vulnerability tasks: function and line-level vulnerability prediction, vulnerability classification, severity estimation, and vulnerability repair. We compare ChatGPT with state-of-the-art language models designed for software vulnerability purposes. Through an empirical assessment employing extensive real-world datasets featuring over 190,000 C/C++ functions, we found that ChatGPT achieves limited performance, trailing behind other language models in vulnerability contexts by a significant margin. The experimental outcomes highlight the challenging nature of vulnerability prediction tasks, requiring domain-specific expertise. Despite ChatGPT's substantial model scale, exceeding that of source code-pre-trained language models (e.g., CodeBERT) by a factor of 14,000, the process of fine-tuning remains imperative for ChatGPT to generalize for vulnerability prediction tasks. We publish the studied dataset, experimental prompts for ChatGPT, and experimental results at \url{https://github.com/awsm-research/ChatGPT4Vul}.
\end{abstract}


\input{sections/introduction}

\input{sections/related_work}

\input{sections/prompting}

\input{sections/experiment}

\section{Conclusion}
\label{sec:conclusion}
In this paper, we empirically evaluate the performance of prompting two versions of ChatGPT (gpt-3.5-turbo and gpt-4) for four common vulnerability tasks: locating vulnerabilities, identifying vulnerability types, estimating severity scores, and suggesting repair patches. We compare the performance of ChatGPT with other pre-trained language models that have significantly smaller model sizes than ChatGPT but have been fine-tuned to perform software vulnerability prediction tasks. Through an assessment encompassing over 190,000 real-world C/C++ functions, ChatGPT yielded the least favorable outcomes across all vulnerability-related tasks, notably struggling to generate accurate patches for the vulnerability repair task. These findings highlight the imperative of possessing security expertise in addressing software vulnerability prediction tasks, a facet not assimilated by ChatGPT during its extensive pre-training phase. Thus, an additional round of fine-tuning stands as a pivotal requirement for ChatGPT to effectively generalize and undertake software vulnerability tasks.
\bibliographystyle{IEEEtranS}
\bibliography{reference}

\end{document}

%% file: sections/introduction.tex
\section{Introduction}
\label{sec:intro}
Software vulnerabilities are weaknesses or flaws in software code that can be exploited by attackers to compromise the security of a system, gain unauthorized access, or cause unintended behavior.
Recently, there have been advancements in employing language models for source code (e.g., CodeBERT, GraphCodeBERT, and CodeT5) to automatically achieve the following tasks: (1) pinpoint vulnerable functions and statements~\cite{fu2022linevul} within source code; (2) recognize vulnerability types to explain detected vulnerabilities~\cite{fu2023aibughunter}; (3) estimate the severity of vulnerabilities~\cite{fu2023aibughunter}; and (4) suggest repair patches~\cite{vrepair,fu2022vulrepair}. In particular, a deep learning-based software security tool named AIBugHunter was proposed in VSCode that achieves promising results for the aforementioned four vulnerability tasks using multiple fine-tuned language models for source code~\cite{fu2023aibughunter}.

On the contrary, large language models (LLMs) like ChatGPT have effectively demonstrated their competence in tasks related to code, such as the simulation of system behavior from provided requirements, the formulation of API specifications, and the discernment of implicit assumptions within code~\cite{white2023chatgpt}.
Leveraging ChatGPT's considerable scale, with 175 billion parameters for gpt-3.5-turbo~\cite{brown2020language} and 1.7 trillion parameters for gpt-4~\cite{openai2023gpt4}, offers the potential for its application in vulnerability-related tasks.
However, to the best of our knowledge, no comprehensive studies have been conducted to evaluate the entire vulnerability workflow, spanning from detecting vulnerabilities and explaining their types to estimating their severity and repair suggestions.

In this paper, we conduct a thorough analysis to assess ChatGPT's ability for the four vulnerability prediction tasks mentioned above. Noteworthy is the fact that ChatGPT's 1.7 trillion parameters surpass the count of parameters in source code-oriented pre-trained language models like CodeBERT and GraphCodeBERT by nearly 14,000 times. Therefore, the prevalent approach to utilizing ChatGPT involves furnishing it with appropriate prompts and task examples, rather than engaging in fine-tuning for these specific downstream tasks. It is important to note that the model parameters of ChatGPT remain proprietary by OpenAI, thereby precluding the possibility of fine-tuning its parameters for vulnerability tasks.

Thus, we compare prompting ChatGPT with other fine-tuned language models specifically designed for source code purposes.
We conduct experiments to compare two versions of ChatGPT (i.e., gpt-3.5-turbo and gpt-4) with four competitive baseline approaches (i.e., AIBugHunter~\cite{fu2023aibughunter}, CodeBERT~\cite{feng2020codebert}, GraphCodeBERT~\cite{guographcodebert}, and VulExplainer~\cite{vulexplainer}) designed for software vulnerability on four different vulnerability tasks.
Through an extensive evaluation of ChatGPT on two vulnerability datasets (i.e., Big-Vul~\cite{fan2020ac} and CVEFixes~\cite{bhandari2021cvefixes}) encompassing over 190,000 C/C++ functions, we answer the following four research questions:

\noindent\textbf{(RQ1) }\rqone\\
\smallsection{Results}
ChatGPT achieves F1-measure of 10\% and 29\% and top-10 accuracy of 25\% and 65\%, which are the lowest compared with other baseline methods.

\noindent\textbf{(RQ2) }\rqtwo\\
\smallsection{Results}
ChatGPT achieves the lowest multiclass accuracy of 13\% and 20\%, which is 45\%-52\% lower than the best baseline.

\noindent\textbf{(RQ3) }\rqthree\\
\smallsection{Results}
ChatGPT gave the most inaccurate severity estimation with the highest mean squared error (MSE) of 5.4 and 5.85 while other baseline methods achieve MSE of 1.8 to 1.86.

\noindent\textbf{(RQ4) }\rqfour\\
\smallsection{Results}
ChatGPT failed to generate any correct repair patches while other baselines correctly repaired 7\%-30\% of vulnerable functions.

\indent \smallsection{\underline{Novelty \& Contributions}}
This paper represents one of the pioneering pilot studies that comprehensively assess ChatGPT's (gpt-3.5-turbo and gpt-4) performance in vulnerability detection, vulnerability type identification, severity estimation, and patch recommendation. In addition, we conduct comparative analyses with other state-of-the-art language models, specifically fine-tuned for software vulnerability-related tasks.


%% file: sections/related_work.tex
\section{Related Work}
\label{sec:related_work}
Recently, researchers have been investigating the applicability of ChatGPT for software vulnerability tasks. Cheshkov~\ea~\cite{cheshkov2023evaluation} investigated the base ChatGPT (gpt-3.5-turbo) performance for vulnerability prediction and classification using 120 samples across five different CWE-IDs. Zhang~\ea~\cite{zhang2023prompt} designed suitable prompts for ChatGPT to enhance its performance for vulnerability prediction. On the other hand, Napoli~\ea~\cite{napoli2023evaluating} investigated ChatGPT's performance for the smart contracts vulnerability correction task. Some previous studies have evaluated ChatGPT's performance for the automated program repair (APR) task where the model was asked to fix general bugs~\cite{pearce2023examining,xia2023keep,sobania2023analysis}. In particular, Pearce~\ea~\cite{pearce2023examining} assessed large language models' performance for the program repair, however, the most advanced two versions of ChatGPT were not included in their experiments.


%% file: sections/prompting.tex
\section{Problem Statement \& Prompt Design}
\label{sec:prompt}
In this section, we introduce the problem statements of the four vulnerability tasks, i.e., (1) function and line-level software vulnerability prediction (SVP), (2) software vulnerability classification (SVC), (3) severity estimation, and (4) automated vulnerability repair (APR). After each problem statement, we illustrate how we design the prompts for ChatGPT to perform the prediction task.

\subsection{Prompt ChatGPT for Software Vulnerability Prediction}
\smallsection{Problem}
We formulate vulnerability prediction as a binary classification task where the model predicts whether the input source code function is vulnerable. For vulnerable functions, we formulate the line-level vulnerability localization task as a ranking problem, where the model ranks vulnerable statements on the top to reduce the manual analysis workload for security analysts.

\begin{figure}[ht]
\includegraphics[width=0.7\linewidth]{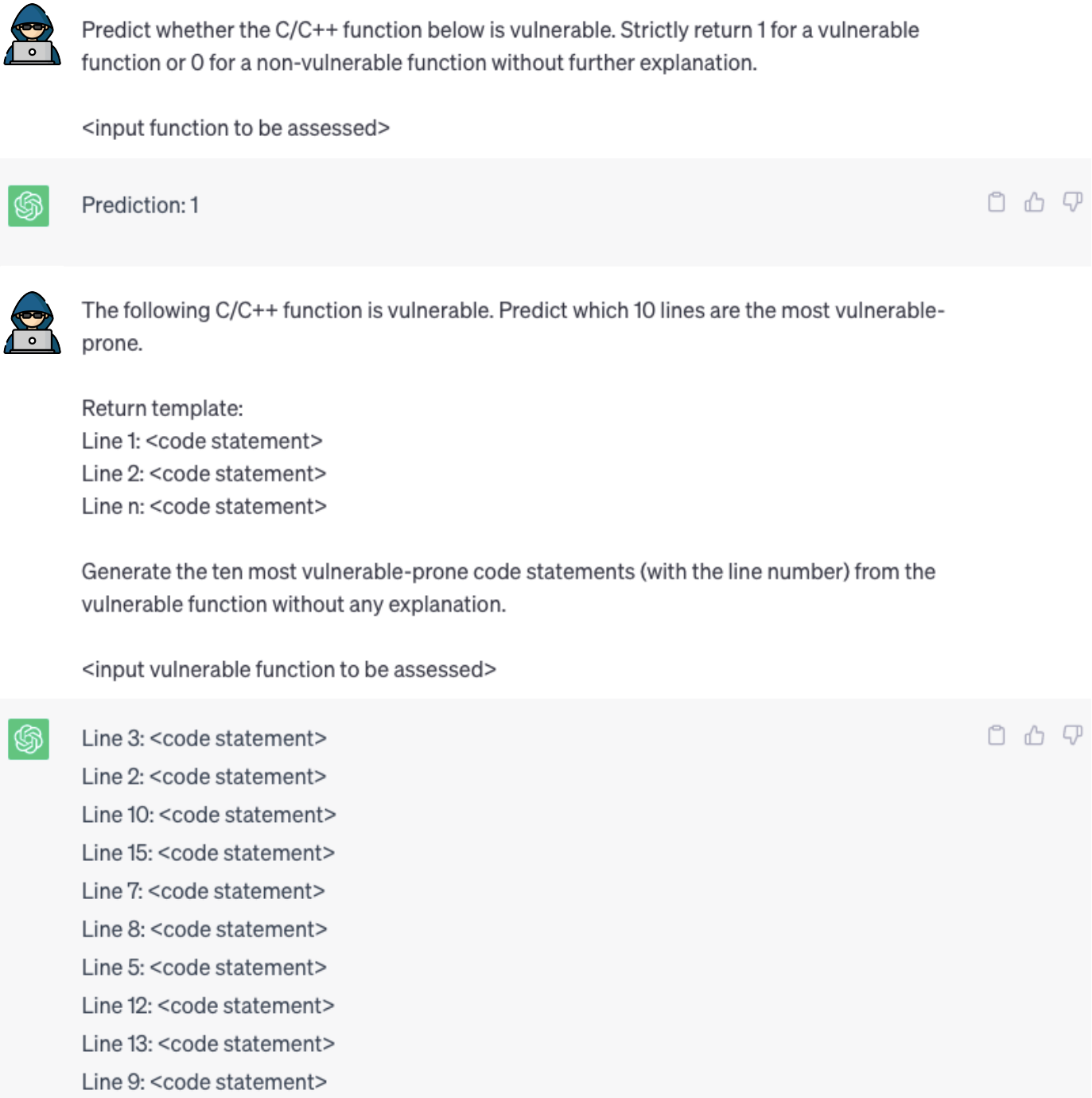}
\centering
\caption{An example prompt for function and line-level vulnerability prediction.}
\label{fig:svp_prompt}
\vspace{-2mm}
\end{figure}

\smallsection{Prompt}
We present example prompts for function and line-level vulnerability prediction in Fig~\ref{fig:svp_prompt}.
In the initial prompt, we provide ChatGPT with a task description focusing on function-level predictions, along with a clear instruction for return.
In the subsequent prompt, we inform ChatGPT that the given function is vulnerable and request it to rank the top 10 most vulnerable-prone statements from the given function.
We provide a return template, anticipating that ChatGPT will generate an output consisting of a line number accompanied by its corresponding code statement as predictions.

\subsection{Prompt ChatGPT for Software Vulnerability Classification}
\smallsection{Problem}
We formulate vulnerability classification as a multiclass classification task where the model identifies a CWE-ID for an input vulnerable function.
Common Weakness Enumeration Identifier (CWE-ID) is a community-developed list of common software weaknesses and vulnerabilities~\cite{cwe}, which allows security professionals to categorize and communicate about security issues in a standardized manner.

\begin{figure}[ht]
\includegraphics[width=0.7\linewidth]{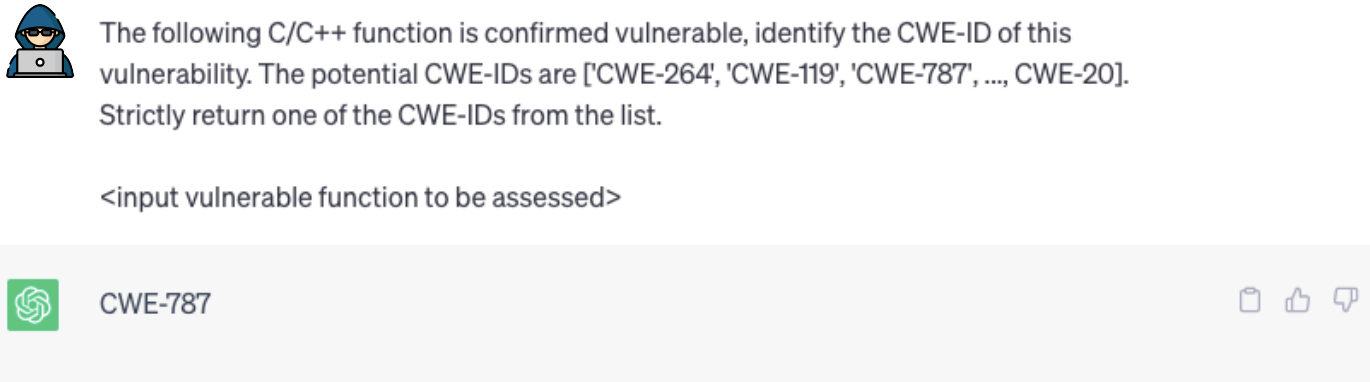}
\centering
\caption{An example prompt for CWE-ID classification.}
\label{fig:svc_prompt}
\end{figure}

\smallsection{Prompt}
We present example prompts for CWE-ID classification in Fig~\ref{fig:svc_prompt}.
In particular, we inform ChatGPT that the input function is vulnerable and request it to identify its corresponding CWE-ID. Additionally, we limit the classification scope by providing a list of potential CWE-IDs to ensure a fair comparison with other fine-tuned models.

\subsection{Prompt ChatGPT for Vulnerability Severity Estimation}
\smallsection{Problem}
We formulate vulnerability severity estimation as a regression task where the model predicts a continuous value based on input vulnerable functions to estimate their severity.
CVSS (Common Vulnerability Scoring System) severity score is a standardized numerical system used to assess the seriousness of security vulnerabilities in software and systems. We use CVSS version 3.1 ranging from 0 to 10.

\begin{figure}[ht]
\includegraphics[width=0.7\linewidth]{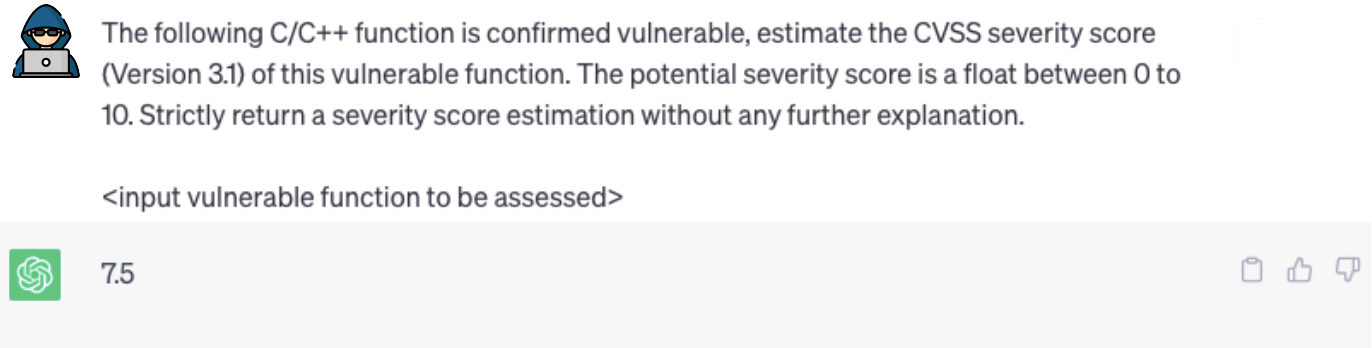}
\centering
\caption{An example prompt for severity estimation.}
\label{fig:sev_prompt}
\end{figure}

\smallsection{Prompt}
We present example prompts for severity estimation in Fig~\ref{fig:sev_prompt}.
We inform ChatGPT that the input function is vulnerable and specify the CVSS version and the output range to make it generate a severity estimation for the given function.

\subsection{Prompt ChatGPT for Automated Vulnerability Repair}
\smallsection{Problem}
We formulate vulnerability repair as a sequence-to-sequence generation task where the model generates corresponding repair patches for input vulnerable functions.

\begin{figure}[ht]
\includegraphics[width=0.7\linewidth]{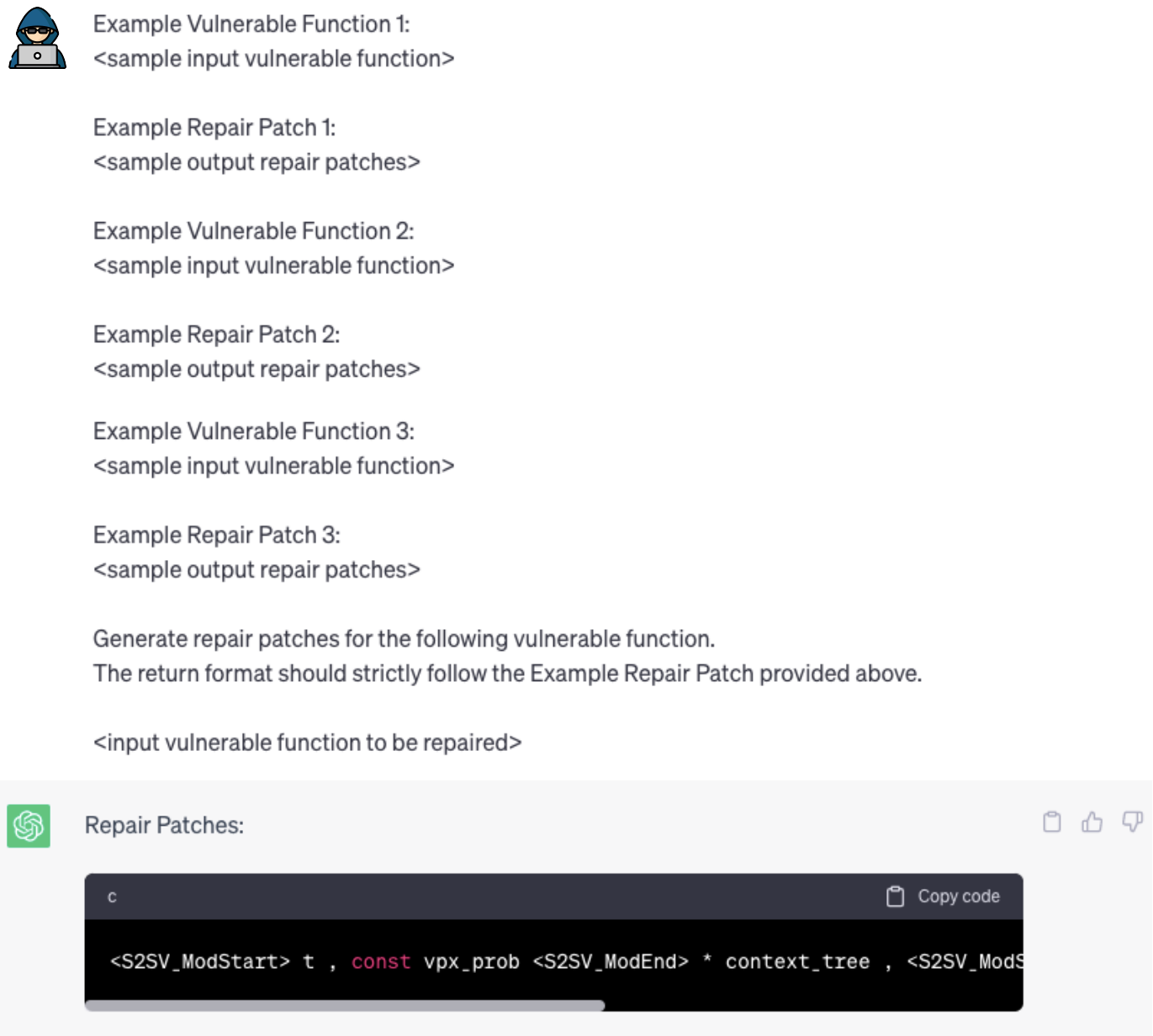}
\centering
\caption{An example prompt for automated vulnerability repair.}
\label{fig:avr_prompt}
\end{figure}

\smallsection{Prompt}
We present example prompts for vulnerability repair in Fig~\ref{fig:avr_prompt}.
Given that model outputs are repair patches designed by Chen~\ea~\cite{vrepair} instead of the complete repaired program, we provide three repair examples in each prompt to make ChatGPT comprehend our repair task. We then request ChatGPT to create repair patches for vulnerable functions using the templates provided in those examples.

%% file: sections/experiment.tex
\section{Experimental Design and Results}
\label{sec:exp}
In this section, we introduce our experimental datasets selected for each vulnerability task followed by the parameter settings and hardware environment used to reproduce the baseline language models fine-tuned for vulnerability tasks. Finally, we present our experimental approach along with the results for each research question.

\subsection{Experimental Datasets}
We use the Big-Vul dataset constructed by Fan~\ea~\cite{fan2020ac} to evaluate vulnerability prediction (RQ1), classification (RQ2), and severity estimation (RQ3). Big-Vul has been widely adopted for software vulnerability tasks~\cite{fu2022linevul,fu2023aibughunter}, which comprises 188k C/C++ functions gathered from 348 Github projects, encompassing 3,754 code vulnerabilities across 91 types. Each vulnerable function is labeled with a CWE-ID and a CVSS severity score. Its data distribution mirrors real-world conditions, with a vulnerable-to-benign function ratio of 1:20. Similar to previous studies~\cite{fu2022linevul,fu2023aibughunter}, we split the data into 80\% for training, 10\% for validation, and 10\% for testing.

For the automated vulnerability repair (RQ4), we leverage the Big-Vul and CVEFixes~\cite{bhandari2021cvefixes} datasets pre-processed by Chen~\ea~\cite{vrepair}. The dataset has been adopted to evaluate vulnerability repair approaches~\cite{vrepair,fu2022vulrepair}, which contains 5.5k pairs of vulnerable functions and their repair patches. Similar to previous studies~\cite{vrepair,fu2022vulrepair}, we split the data into 70\% for training, 10\% for validation, and 20\% for testing.

\subsection{Parameter Settings and Execution Environment}
We replicate the baseline methods using the original authors' specified parameter settings, running experiments on a Linux machine equipped with an AMD Ryzen 9 5950X processor, 64 GB RAM, and an NVIDIA RTX 3090 GPU. The ChatGPT prompting was completed via paid API access provided by OpenAI~\cite{chatgpt}.

\subsection{Experimental Results}
\subsection*{\textbf{(RQ1) \rqone}}
\smallsection{Approach}
To answer this RQ, we focus on the function and line-level vulnerability predictions and compare ChatGPT (i.e., gpt-3.5-turbo and gpt-4) with three other fine-tuned baseline language models as follows:
\begin{enumerate}
    \item AIBugHunter: A recently proposed deep learning-based software security tool that utilizes fine-tuned language models~\cite{fu2022linevul,fu2022vulrepair} to perform vulnerability prediction, classification, severity estimation, and repair~\cite{fu2023aibughunter}.
    \item CodeBERT: A language model originally pre-trained for tasks related to source code, CodeBERT underwent pre-training using the Codesearchnet dataset~\cite{husain2019codesearchnet}, encompassing various programming languages. CodeBERT has demonstrated its capability to effectively perform tasks associated with source code~\cite{feng2020codebert}. 
    \item GraphCodeBERT: A language model that was also pre-trained on the Codesearchnet dataset to perform source code-related tasks. Notably, when forming the input for the model, GraphCodeBERT considers the data flow graph in addition to source code tokens~\cite{guographcodebert}.
\end{enumerate}
Similar to the previous study~\cite{fu2022linevul}, we report F1-measure, precision, and recall to evaluate function-level performance. For line-level performance, we report top-10 accuracy that measures the percentage of vulnerable functions where at least one actual vulnerable line appears in the model's top-10 ranking. This metric has been previously used to evaluate the prediction of line-level vulnerability prediction~\cite{fu2022linevul}.

\begin{figure*}[t]
\includegraphics[width=0.8\linewidth]{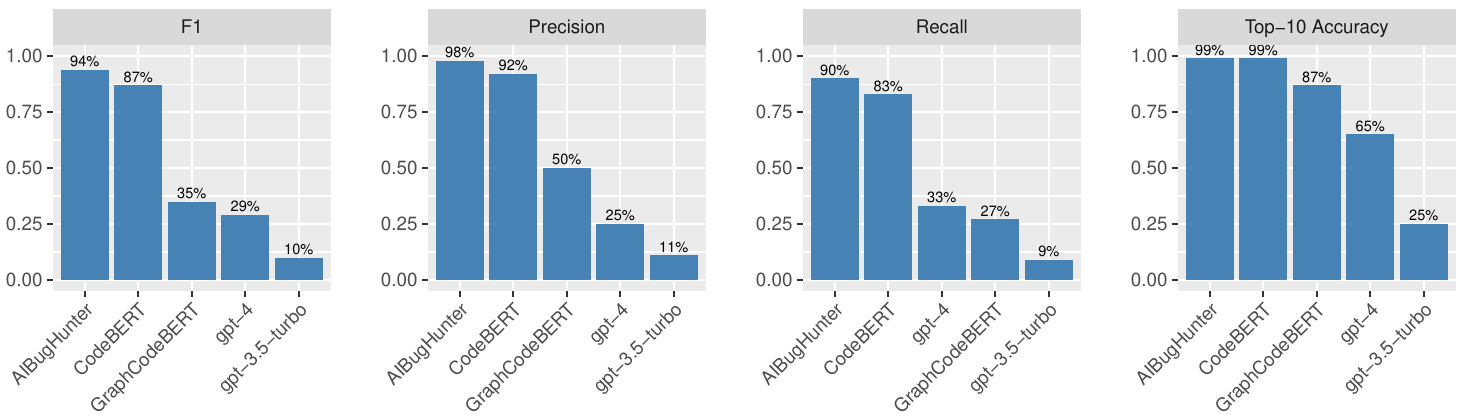}
\centering
\caption{(RQ1) The experimental results of function-level and line-level vulnerability prediction. ($\nearrow$) For all metrics, higher = better.}
\label{fig:rq1_result}
\end{figure*}
\smallsection{Result}
Fig~\ref{fig:rq1_result} presents the experimental results of function and line-level vulnerability prediction.
\textbf{ChatGPT failed to accurately predict at the function level with an F1-measure of 10\% and top-10 accuracy of 25\% at the statement level.}
The leading-edge gpt-4 with 1.7 trillion parameters achieves an F1-measure of 29\% along with a top-10 accuracy of 65\%.
In contrast, the fine-tuned AIBugHunter achieves an F1-measure of 94\% along with a top-10 accuracy of 99\% with only 120 million parameters.
Despite gpt-4's extensive model size and pre-training data, it faced challenges in generalizing the vulnerability prediction task without undergoing fine-tuning.
These results highlight that the vulnerability prediction task requires models to learn domain-specific knowledge (e.g., vulnerability patterns) and fine-tuning is still required for large language models despite their significant model size.
\subsection*{\textbf{(RQ2) \rqtwo}}
\smallsection{Approach}
To answer this RQ, we focus on the vulnerability classification task where we aim to identify CWE-IDs for vulnerable functions. We compare ChatGPT (i.e., gpt-3.5-turbo and gpt-4) with three fine-tuned baseline language models introduced in RQ1. Additionally, we include VulExplainer~\cite{vulexplainer} which leverages language models with a distillation framework to mitigate the data imbalances in the CWE-ID classification task. Similar to previous studies~\cite{fu2023aibughunter,vulexplainer}, we use the multiclass accuracy measure to evaluate the performance of each method.

\begin{figure}[t]
\includegraphics[width=0.4\linewidth]{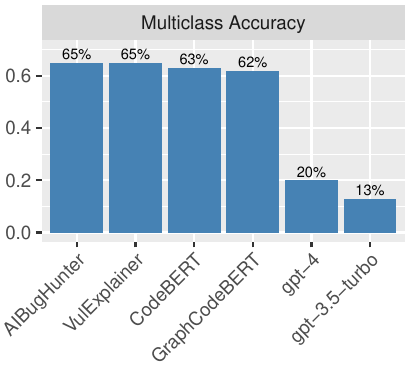}
\centering
\caption{(RQ2) The experimental results of vulnerability type (i.e., CWE-ID) classification. ($\nearrow$) Higher Multiclass Accuracy = better.}
\label{fig:rq2_result}
\end{figure}
\smallsection{Result}
Fig~\ref{fig:rq2_result} presents the experimental results of CWE-ID classification.
\textbf{The accuracy of ChatGPT in correctly identifying CWE-IDs for vulnerable functions is limited, standing at a mere 13\%.}
The gpt-3.5-turbo and gpt-4 achieve 13\%-20\% accuracy while the fine-tuned language model baselines achieve 62\%-65\%.
These findings suggest that the accurate identification of CWE-ID for a vulnerable function requires the model to learn to map specific patterns (e.g., buffer overflow) in vulnerable functions to a CWE-ID. However, ChatGPT has not adequately acquired such knowledge during the pre-training phase of ChatGPT so a fine-tuning stage is still required to boost its performance.

\subsection*{\textbf{(RQ3) \rqthree}}
\smallsection{Approach}
To answer this RQ, we focus on predicting the CVSS score of vulnerable functions. We compare ChatGPT (gpt-3.5-turbo and gpt-4) with the three baselines introduced in RQ1, where CodeBERT has been shown to be effective for severity estimation~\cite{fu2023aibughunter}. Similar to the previous study~\cite{fu2023aibughunter}, we use Mean Squared Error (MSE) and Mean Absolute Error (MAE) to assess the performance of each method.

\begin{figure}[t]
\includegraphics[width=0.8\linewidth]{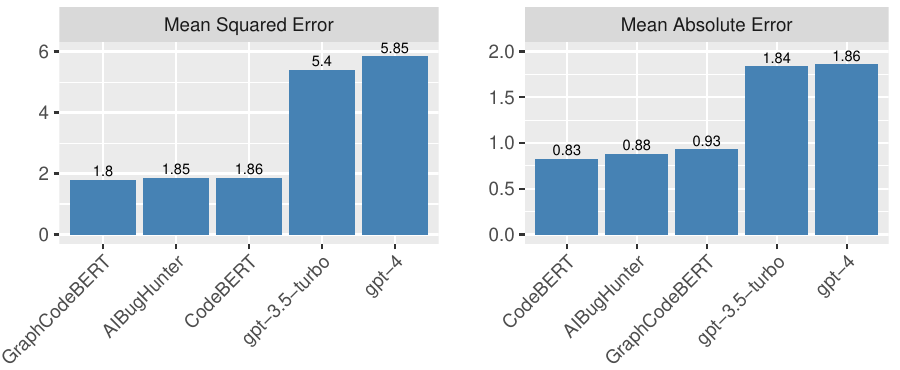}
\centering
\caption{(RQ3) The experimental results of vulnerability severity estimation. ($\searrow$) Lower MSE, MAE = better.}
\label{fig:rq3_result}
\end{figure}
\smallsection{Result}
Fig~\ref{fig:rq3_result} presents the experimental results of the severity score estimation.
\textbf{ChatGPT failed to accurately estimate the CVSS severity score, resulting in an MSE of 5.4 and an MAE of 1.84.}
The gpt-3.5-turbo and gpt-4 have MSE of 5.4 and 5.85 while the fine-tuned language model baselines achieve 1.8-1.86.
Similar to vulnerability prediction and classification tasks, accurately estimating severity scores also demands software security expertise that ChatGPT has not acquired during its extensive pre-training phase, hindering its ability to provide accurate predictions in this context.

\subsection*{\textbf{(RQ4) \rqfour}}
\smallsection{Approach}
To answer this RQ, we focus on generating vulnerability repair patches for vulnerable functions. We compare ChatGPT (i.e., gpt-3.5-turbo and gpt-4) with the three baseline methods introduced in RQ1. Notably, AIBugHunter leverages the VulRepair~\cite{fu2022vulrepair} model for repair patches generation, which achieves state-of-the-art results in the vulnerability repair problem. Similar to previous studies~\cite{vrepair,fu2022vulrepair}, we use the percentage of perfect prediction (\%PP) measure to assess the performance of each method. Only if the generated repair patches are identical to the ground-truth patches, we count it as a correct prediction. The \%PP is computed as $\frac{\text{total correct predictions}}{\text{total testing samples}}$. We use greedy decoding to return one repair candidate for fine-tuned language models and ensure a fair comparison with ChatGPT. Furthermore, we incorporate BLEU and METEOR scores to evaluate the degree of similarity between the patches produced by the model and the actual patches.

\begin{figure}[t]
\includegraphics[width=\linewidth]{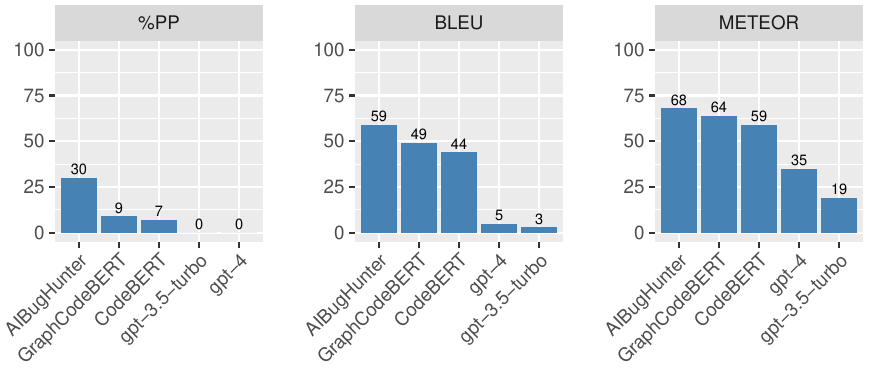}
\centering
\caption{(RQ4) The experimental results of automated vulnerability repair. ($\nearrow$) Higher \%PP, BLEU, METEOR = better.}
\label{fig:rq4_result}
\end{figure}
\smallsection{Result}
Fig~\ref{fig:rq4_result} presents the experimental results of the vulnerability repair.
\textbf{ChatGPT failed to generate correct repair patches for all of the vulnerable functions in our testing data.}
In contrast, the fine-tuned language model baselines can correctly repair 7\%-30\% of the testing function.
The BLEU and METEOR scores further demonstrate that repair patches generated using baseline methods exhibit greater proximity to the true patches compared to those generated through ChatGPT methods.
These results indicate that vulnerability repair is a more challenging task compared with other vulnerability prediction tasks, where ChatGPT struggle to generate correct repairs for vulnerable functions. Thus, a fine-tuning step using domain-specific data is crucial for ChatGPT to generalize its ability for the vulnerability repair task.